\documentclass[11pt]{article}
\usepackage{amsmath,amssymb,cite,epsfig}
\catcode`\@=11

\long\def\@makefntext#1{\parindent 0cm\noindent
\hbox to 1em{\hss$^{\@thefnmark}$}#1}

\numberwithin{equation}{section}

\begin{document}
\begin{titlepage}
\begin{flushright}
UCD-2003-07\\
gr-qc/0306054\\
June 2003\\
\end{flushright}
\bigskip

\begin{center}
{\Large{\bf Phase Transitions and Critical Behavior\\[1ex] for Charged Black Holes}}
\bigskip

S. Carlip\footnote{carlip@dirac.ucdavis.edu} and 
S. Vaidya\footnote{vaidya@dirac.ucdavis.edu} \\
{\it Department of Physics, \\
University of California, Davis, CA 95616, USA} \\
\end{center}

\vspace{.5in}
\begin{center}
{\large\bf Abstract}
\end{center}
\begin{center}
\begin{minipage}{4.75in}
{\small
We investigate the thermodynamics of a four-dimensional charged 
black hole in a finite cavity in asymptotically flat and asymptotically 
de Sitter space.  In each case, we find a Hawking-Page-like phase 
transition between a black hole and a thermal gas very much like 
the known transition in asymptotically anti-de Sitter space.  For 
a ``supercooled'' black hole---a thermodynamically unstable black 
hole below the critical temperature for the Hawking-Page
phase transition---the phase diagram has a line of first-order phase 
transitions that terminates in a second order point.  For the 
asymptotically flat case, we calculate the critical exponents at 
the second order phase transition and find that they exactly match 
the known results for a charged black hole in anti-de Sitter space.  We 
find strong evidence for similar phase transitions for the de Sitter black 
hole as well.  Thus many of the thermodynamic features of charged
anti-de Sitter black holes do not really depend on asymptotically
anti-de Sitter boundary conditions; the thermodynamics of charged 
black holes is surprisingly universal.}
\end{minipage}
\end{center}
\end{titlepage}
\addtocounter{footnote}{-2}

\section{Introduction}

It has been known for some time that a black hole in asymptotically
anti-de Sitter space can undergo a phase transition, ``evaporating''
at a critical temperature into a hot gas \cite{hp}.  This Hawking-Page
phase transition has recently garnered new attention with the
realization that it can be used to test the AdS/CFT correspondence
by identifying a similar structure in a dual conformal field theory
\cite{wit}.  A dual description may exist not only for this particular 
transition, but for a number of finer details of the phase structure of 
the asymptotically AdS black hole as well \cite{cejm,cejm2}.  The purpose 
of this paper is to show that this phase structure is not unique to
the asymptotically AdS black hole, but occurs universally for suitably 
stabilized black holes in asymptotically flat and asymptotically de 
Sitter space.

A black hole in anti-de Sitter space can be thermodynamically stable.
In asymptotically flat or de Sitter space, this is no longer the case: an
isolated black hole radiates away its energy in the form of Hawking
radiation. To understand equilibrium black hole thermodynamics, one 
must therefore work with ensembles that include not just the black hole,
but also its environment \cite{York}. As self-gravitating systems are 
spatially inhomogeneous, any specification of such ensembles requires 
not just thermodynamic quantities of interest, but also the place at 
which they take the specified values. Having set up an appropriate 
thermodynamic ensemble, one can then proceed to ask interesting 
questions regarding stability and phase structure.

Such an analysis was done for the case of charged black holes in
asymptotically flat space by Braden, Brown, Whiting, and York \cite{bbwy}. 
Following a brief summary of their analysis, we will build on those results 
to study the phase structure of charged black holes. We will also apply a
similar method to charged black holes in de Sitter space.  This work grew
out of an effort to understand the effect of varying ``constants'' on black 
hole thermodynamics \cite{cv}, but it may have broader significance.  In 
particular, the universality of the phase structure may have interesting
implications for holography in asymptotically flat and de Sitter space.

The Euclidean action \cite{gw} for metric $g$ and electromagnetic
field $A_\mu$ over a region $M$ with boundary $\partial M$ takes the
form
\begin{equation}
I = -\frac{1}{16 \pi} \int_M d^4 x \sqrt{g}\, [(R- 2\Lambda) -F^2] +
\frac{1}{8 \pi} \int_{\partial M} d^3x \sqrt{\gamma}\, (K - K_0),
\label{action}
\end{equation} 
where the cosmological constant $\Lambda$ is zero for flat space and
positive for de Sitter space. The geometric and electromagnetic data
specified at the boundary $\partial M$ fixes for us the thermodynamic
data for this gravitational system thought of as a statistical
ensemble.  We take a spherically symmetric ansatz for the metric of a 
black hole spacetime,
\begin{equation} 
ds^2 = f^2 d\tau^2 + \alpha^2 dy^2 + r^2 d\Omega^2 ,
\label{metric}
\end{equation}
where $f, \alpha$ and $r$ are functions of the radial coordinate $y
\in [0,1]$.  The reduced action may be calculated from (\ref{action})
provided certain conditions are taken into account:
\begin{enumerate}
\item The horizon of the black hole is located at $y=0$ (i.e. $r(0)=r_+$), 
and we restrict to the class of metrics that are ``regular.''  In other words, 
the geometry of the spacetime near the origin of the $y$-$\tau$ plane looks
like the plane ${\mathbb R}^2$, so the horizon is nondegenerate.
\item The boundary at $y=1$ has a topology $S^1 \times S^2$ with the
two-sphere having an area $4\pi r_B^2$. Heat can flow in either
direction through this boundary in such a manner as to keep the
temperature $\beta^{-1}$ fixed. The inverse temperature is simply the
proper length of the circle $S^1$ of the boundary: $\beta=2 \pi f(1)$
(where we take $\tau$ to have period $2\pi$).
\end{enumerate}
We also need to specify appropriate electromagnetic data at the
boundary $y=1$. For the grand canonical ensemble, we specify 
the gauge potential $A_\tau(1)$ (more precisely the difference of potential 
between $y=0$ and $y=1$). For the canonical ensemble, we must add a 
boundary term to the usual Maxwell action and hold the electromagnetic 
field fixed at the boundary. This is equivalent to keeping the total charge 
inside the cavity fixed.

Using this analysis, we first demonstrate in section 2 that black holes 
in asymptotically flat space show evidence of a Hawking-Page phase 
transition to a thermal gas.  While the transition temperature depends 
on the charge, the variation is surprisingly small.  In the absence of a 
well-understood comparison action for a charged gas, conclusions 
from (fixed-charge) canonical ensemble are somewhat ambiguous; we 
therefore also check for this transition in the grand canonical ensemble,
and confirm its existence.

We next investigate the unstable ``supercooled'' region of the phase 
diagram, and show that black holes in this region, like those in anti-de 
Sitter space \cite{cejm,cejm2},  exhibit a first order phase transition: 
for a given (small) value of charge, the entropy of the system changes
discontinuously as the temperature is varied. As the charge increases,
the ``height'' of this discontinuity decreases, and the line of first
order phase transitions terminates at a critical point. We
show that this critical point is the location of a second order phase
transition, and we calculate the critical exponents associated with
it.  Remarkably, the critical exponents are the same as those found
by Chamblin et al.\ \cite{cejm,cejm2} for charged black holes in 
anti-de Sitter space, hinting that these two systems belong to the same 
universality class.

In section 3, we apply the reduced action technique to charged black
holes in de Sitter space.  We again find evidence for a Hawking-Page 
phase transition.  For sufficiently small cosmological constant,
the supercooled region again displays a line of first order phase
transitions terminating at a critical point. This picture no longer
remains valid as $\Lambda$ increases. In particular, it seems that for
large values of $\Lambda$, there is no second order transition:
the location of the would-be transition is pushed beyond the allowed
boundaries of the cavity.

The reduced action for the de Sitter case is more complicated, and we
cannot find the critical exponents of the second order phase
transition (when it exists) analytically. We can study two special
cases: the uncharged black hole for any value of $\Lambda$ and the
charged black hole for small $\Lambda$.  We again find the same
structure, with critical exponents that are unchanged by the presence 
of a small cosmological constant.

Section 4 is devoted to discussion our results.

\section{Charged black hole in asymptotically flat space}

Following Brown et al.\ \cite{bbwy}, we first insert the ansatz 
(\ref{metric}) into the action (\ref{action}), and find that the 
Hamiltonian constraint can be solved, yielding
\begin{equation} 
V(r) \equiv \left(\frac{r'}{\alpha}\right)^2 = 1-\frac{C}{r} +
\frac{e^2}{r^2}.
\end{equation} 
For a nondegenerate horizon, $C=r_+ + e^2/r_+$, with $e^2<r_+^2$. 
The reduced action for the canonical ensemble then takes the form
\begin{equation} 
I_C = \beta r_B \left( 1-\sqrt{\left(1-\frac{r_+}{r_B}\right)
  \left(1-\frac{e^2}{r_+ r_B}\right)}\right) - \pi r_+^2.  
\end{equation} 
Extremizing the action with respect to $r_+$ gives us
\begin{equation} 
\beta = 4 \pi r_+\left(1-\frac{e^2}{r_+^2}\right)^{-1}
\left(1-\frac{r_+}{r_B} \right)^{1/2} \left(1-\frac{e^2}{r_+ r_B}
\right)^{1/2},
\label{bvsr}
\end{equation} 
which should be viewed as an equation for the unknown $r_+$ in terms of the
fixed charge, boundary radius, and temperature of the canonical ensemble.
Each extremum gives a saddle point contribution to the Euclidean path 
integral for the canonical partition function $Z[\beta]$, with a Helmholtz
free energy and entropy
\begin{eqnarray}
F &=& -\frac{1}{\beta}\ln Z[\beta] =  \frac{1}{\beta}I_C[\beta,r_+(\beta)] \nonumber\\
S &=&  - \left( \beta\frac{\partial\ }{\partial\beta} - 1\right) \ln Z[\beta] 
  = \pi r_+^2.
\label{Entropy}
\end{eqnarray}

Equation (\ref{bvsr}) can be rewritten as the seventh order algebraic equation
\begin{equation} 
x^5(x-1)(x-q^2) + b^2 (x^2 - q^2)^2 = 0
\label{seventh}
\end{equation} 
with $x=r_+/r_B,q=e/r_B$ and $b=\beta/4\pi r_B$.
We are interested only on those solutions that are real positive and satisfy 
$q^2 < x^2$. Of these positive solutions, we initially choose the one that 
minimizes the free energy $F[\beta]$, since this is the thermodynamically 
stable state in the canonical ensemble. 

\subsection{The Hawking-Page transition \label{HP1}}

\begin{figure}
\centerline{%
\epsfig{figure=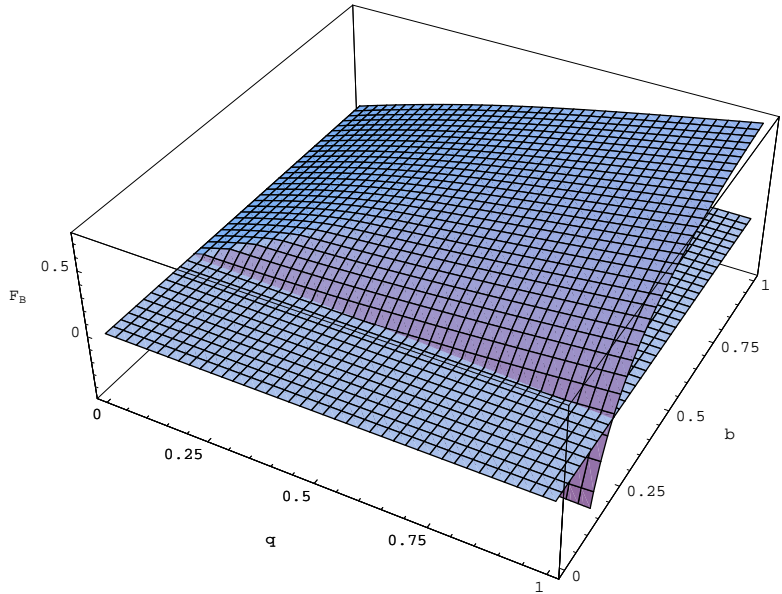,height=3.2in}}
\caption{Free energy of a charged black hole in asymptotically flat space as a
function of inverse temperature $b$ and charge $q$.}
\label{fig:evsq1}
\end{figure}

The plot of free energy as a function of charge $q$ and inverse temperature
$\beta$ is shown in Figure \ref{fig:evsq1}.  For reference, we have also plotted 
the plane $F=0$.  In the original paper of Hawking and Page \cite{hp}, the
phase transition to a hot gas of particles was determined by comparing the
black hole free energy with that of a real thermal gas, but subsequent work 
(for example, \cite{wit}) has often used ``hot empty AdS'' as a reference.  This
alternative is not available for a charged black hole in the canonical ensemble,
except at $q=0$, since there is no ``hot empty charged space''; one cannot
avoid the intricacies of a real charged gas.

\begin{figure}
\centerline{%
\epsfig{figure=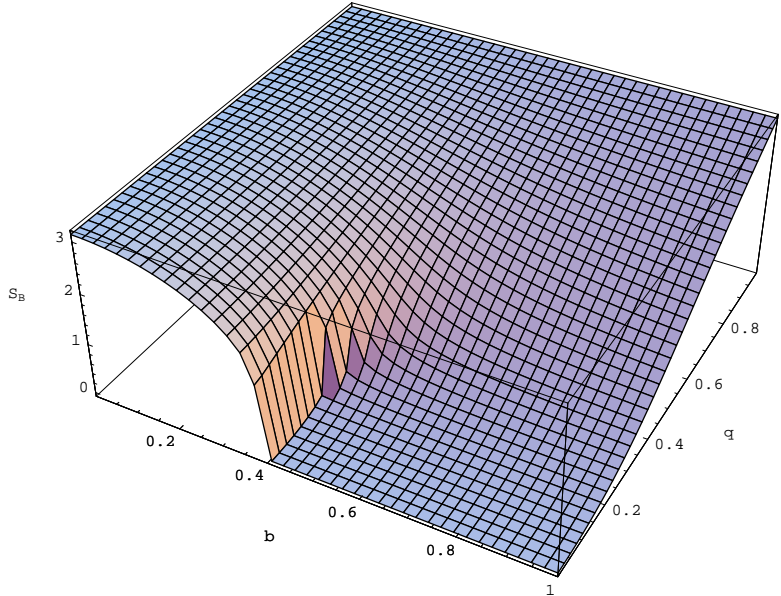,height=3.2in}}
\caption{Entropy of a charged black hole in asymptotically flat space as a
function of inverse temperature $b$ and charge $q$.}
\label{fig:entropy}
\end{figure}

Although there has been a bit of work on the thermodynamics of self-gravitating 
gases of particles \cite{bilic,bilic2,bilic3}, the subject remains rather poorly understood. 
For low temperature, Refs.\ \cite{bilic,bilic2,bilic3} confirm the natural 
guess that the free energy is approximately $Nm$, where $N$ is the 
number of gas particles of mass $m$ and charge $q_0$, while the charge 
is $e=Nq_0$.\footnote{For Newtonian results, see also \cite{thir}.}
For real elementary particles, $q_0\gg m$ in geometrized units---for an
electron, for instance, $q_0/m \sim 10^{21}$---so $F\ll e$, even for a 
near-extremal black hole.  An approximation $F=0$ for a thermal gas 
thus seems reasonable, and the intersection of the $F=0$ plane with the 
free energy in Figure \ref{fig:evsq1} should give the rough location of 
a Hawking-Page-type phase transition.  It is straightforward to check that 
above the transition temperature, the black hole is thermodynamically 
stable, i.e., $C_V = -b(\partial S/\partial b)$ is positive everywhere.  Figure 
\ref{fig:entropy}, for example, plots the entropy against $q$ and $\beta$;
the sign of $C_V$ is determined by the slope.

In the absence of a better understanding of the thermodynamics of hot
self-gravitating gases, this argument is admittedly not completely 
convincing.  The ``competing'' configurations are, after all, not just
dilute gases, but stars.  To check more carefully for a Hawking-Page 
transition, we have therefore followed the suggestion of Ref.\
\cite{cald} and looked at the grand canonical ensemble, in which the 
boundary potential $\phi$ is fixed instead of the charge.  This ensemble 
has the nice feature that it {\em does\/} allow a comparison with ``hot 
empty space'': while an empty cavity cannot contain charge, its walls 
can be held at a constant potential.  

\begin{figure}
\centerline{%
\epsfig{figure=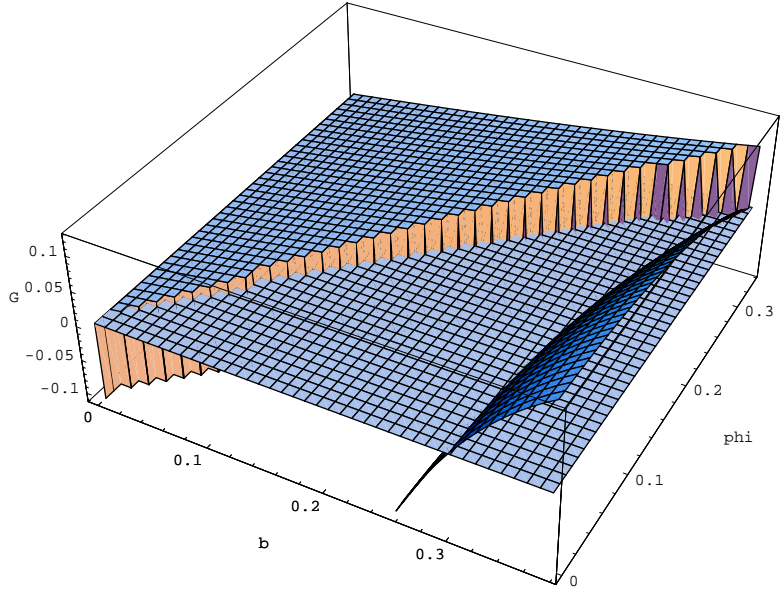,height=3.2in}}
\caption{Gibbs free energy of a charged black hole in asymptotically flat space as a
function of inverse temperature $b$ and potential $\phi$.}
\label{fig:gce1}
\end{figure}

Figure \ref{fig:gce1} shows the Gibbs 
free energy of the grand canonical ensemble, with the plane $G=0$ superimposed.  
The intersection shows the location of the Hawking-Page phase transition to 
hot empty space.\footnote{The Gibbs free energy becomes positive again at
very high temperatures.  This occurs when the thermodynamically stable black 
hole becomes larger than the cavity radius $r_B$, leaving only an unstable 
(negative specific heat) configuration.}   The transition depends only weakly 
on the potential.  There is one important difference between the charged and 
uncharged cases, however.  For an uncharged black hole, there is a further 
critical temperature beyond which the action has no extremum, even with 
positive free energy.  Thus even a ``supercooled'' black hole cannot exist below 
this temperature.  For a charged black hole, no such temperature exists; a 
locally stable black hole can exist for any value of $\beta$, essentially because 
a near-extremal black hole can have an arbitrarily low temperature.

\subsection{Supercooled black holes and critical exponents \label{Unch}}

Let us now return to the canonical ensemble, and consider the portion of Figure 
\ref{fig:evsq1} for which the black hole free energy is positive.  While a black 
hole in this region is globally unstable, it is locally stable---the specific heat is 
positive, as is evident from Figure \ref{fig:entropy}.  One can therefore 
produce a ``supercooled'' black hole in this region, and, indeed, the 
analysis of fluctuations for uncharged black holes suggests that such a
state may have a very long lifetime \cite{katz,paren}.  Figure \ref{fig:evsq1} 
now suggests that there is a line of first order phase transitions in this region
that terminates at a critical point in the $(q,\beta)$-plane. We will show 
that this critical point is the location of a second order phase transition.

The simplest way to see this is to rewrite (\ref{bvsr}) as
\begin{equation} 
b(x,q)=\frac{x(1-x)^{1/2} (1-q^2/x)^{1/2}}{1-q^2/x^2} .
\end{equation} 
In general, $b(x,q)$ has may have extrema (both maxima and minima) as a
function of $x$, with locations  
\begin{equation} 
\frac{\partial b}{\partial x} = \frac{x(q^4 (5-6x)+(2-3x)x^3 + 2q^2x
  (-3 +3x +x^2))}{2 (q^2 -x^2)^2 \sqrt{1-x} \sqrt{1-q^2/x}} = 0 .
\end{equation} 
The numerator is a polynomial in $x$ (with functions of $q$ as
coefficients), with a discriminant 
\begin{equation}   
\Delta(q) = \frac{6400}{243}q^4 (1-q^2)^4 (1- 18q^2 + q^4) .
\end{equation} 
This discriminant determines the condition for the maximum and the 
minimum to coincide: they do so for $\Delta(q)=0$, i.e., for $q_c = -2+\sqrt{5}$.  
The corresponding critical values for the horizon radius and temperature
are $x_c = 5-2\sqrt{5}$ and $b_c = \frac{5}{2} (17+38/\sqrt{5})^{-1/2}$. 
Thus $(x_c,b_c,q_c) \simeq (0.528,0.429,0.236)$. Expanding in the neighborhood 
of the critical point, we obtain
\begin{equation}  
b-b_c = -\frac{1}{8} (425 + 195 \sqrt{5})^{1/2} (x-x_c)^3 + \cdots
\simeq -3.644 (x-x_c)^3 + \cdots .
\end{equation} 
To see the dependence on $x$ near $q_c$, we rewrite (\ref{bvsr}) as 
\begin{equation} 
q(x,b)=\frac{(2b^2 x^2 - x^5 + x^6 - (1-x)x^3
  \sqrt{4b^2+x^4})^{1/2}}{\sqrt{2}b} 
\end{equation} 
and expand in the neighborhood of $(x_c,b_c)$:
\begin{equation} 
q - q_c = \frac{1}{40}(105 + 47\sqrt{5})(x-x_c)^3 + \cdots \simeq
5.252 (x-x_c)^3 + \cdots .
\end{equation}  
Inverting these relations and writing in terms of the entropy $S$, we find
\begin{eqnarray} 
S-S_c &=&  1.226 (\tau - \tau_c)^{1/3} + \cdots, \\
S-S_c &=& 4 1.908(q-q_c)^{1/3} + \cdots, 
\end{eqnarray} 
where $\tau = r_B k T,\tau_c=1/b_c$ and $S_c=\pi x_c^2$.
The specific heat $C_v = T(\partial S/\partial T)=\tau(\partial
S/\partial \tau)$ is
\begin{equation} 
C_v \simeq 0.409 (\tau - \tau_c)^{-2/3}
\end{equation} 
which means that the critical exponent $\alpha$ is $-2/3$.  Remarkably, 
this is also the critical behavior of the charged black hole in asymptotically 
anti-de Sitter space \cite{cejm,cejm2,wu}.  There, as in the asymptotically
flat case discussed here, a line of first order phase transitions in the
``supercooled'' region ends in a second order transition, and although 
the AdS partition function has a completely different functional form,
the critical exponents are identical.

The case of uncharged black hole is even simpler.  Equation (\ref{seventh})
reduces to the cubic $x^3 -x^2 +b^2=0$, with critical values $(x_c,b_c)=
(2/3,2/(3\sqrt{3}))$.  This critical point does not signal a second order phase 
transition \cite{bbwy}: in contrast to the case of even arbitrarily small charge,
the action for the uncharged black hole has no extremum beyond this point.
As discussed at the end of section \ref{HP1}, this point instead describes a
transition between even a supercooled black hole and a thermal gas.  Near  
this point,
\begin{equation} 
S-S_c \sim (\tau - \tau_c)^{1/2}
\end{equation} 
indicating that the specific heat diverges like $C_v \sim (\tau-\tau_c)^{-1/2}$. 
This is that same critical behavior seen in non-spinning $D3$-branes 
\cite{gubser,caisoh}.

\section{Charged black holes in de Sitter space}

Conceptually, the method of analyzing the thermodynamic stability of
black holes in de Sitter space is the same as in flat space.  
Lorentzian de Sitter space has infinite spatial extent, but in static
coordinates there is a cosmological horizon. The corresponding
Euclidean section of the metric has finite volume, with topology of a
four-sphere with a radius equal to the cosmological horizon. 

For a charged
black hole in de Sitter space, the spacetime has, generically, three
horizons: an inner and an outer horizon for the black hole, and the
cosmological horizon.
Any discussion of equilibrium black hole thermodynamics requires one
to specify thermodynamic quantities at a boundary that is inside the
cosmological horizon.  In this case, the entropy of the equilibrium 
configuration depends on three parameters: temperature, charge, 
and the cosmological constant.  Again using the spherically ansatz 
(\ref{metric}) in the action and solving the Hamiltonian constraint, 
we find
\begin{equation} 
V(r)=\left(\frac{r'}{\alpha}\right)^2 = 1 +\frac{e^2}{r^2} - \frac{\Lambda
  r^2}{3} - \frac{C}{r}.
\end{equation} 
The integration constant $C$ is often identified as the mass parameter
$2 M$.

Our first task is to determine the allowed values of the parameters.
For the path integral to be well-defined, we need to restrict to those
values of $e, M, \Lambda$ and $r$ for which $V(r)$ is
nonnegative. Event horizons are located at those values of $r$ for
which $V(r)=0$.  To locate these, it is useful to define
dimensionless quantities $\rho \equiv r \sqrt{\Lambda}, g \equiv e
\sqrt{\Lambda}$ and $\mu \equiv M \sqrt{\Lambda}$, so  
\begin{equation} 
V=-\frac{1}{3 \rho^2}Q(\rho;\mu,g^2)=-\frac{1}{3\rho^2}(\rho^4 -
3\rho^2 +6 \mu \rho -3g^2)
\end{equation} 
The quartic polynomial $Q$ has four real roots: one negative root,
and three positive roots that we denote, in ascending order,  
$\rho_{-}(\mu, g^2)$, $\rho_{+}(\mu, g^2)$, and $\rho_c(\mu, g^2)$. 
The region outside the outer horizon of the black hole is thus 
$\rho_{+}(\mu,g^2)/\sqrt{\Lambda} \leq r \leq \rho_c(\mu,g^2)/
\sqrt{\Lambda}$. The condition for two (or more) of these roots to 
coincide \cite{romans} is given by the vanishing of the discriminant 
of $Q(\rho,\mu,g^2)$:
\begin{equation} 
\Delta(\mu,g^2)=(1-4g^2)^3 - (1+12 g^2 -18 \mu^2)^2 = 0.
\end{equation} 
The the allowed values of mass and charge parameter in the
$(\mu,g)$-plane are thus given by the interior of the region 
bounded by the curves
\begin{eqnarray} 
\mu_{min}(g) &=& \sqrt{\frac{1+12g^2 - \sqrt{(1-4g^2)^3}}{18}}, \\
\mu_{max}(g) &=& \sqrt{\frac{1+12g^2+\sqrt{(1-4g^2)^3}}{18}} ,
\end{eqnarray} 
which intersect at the point $(\mu,g)=(\sqrt{2}/3,1/2)$.
\begin{figure}
\centerline{\epsfig{figure=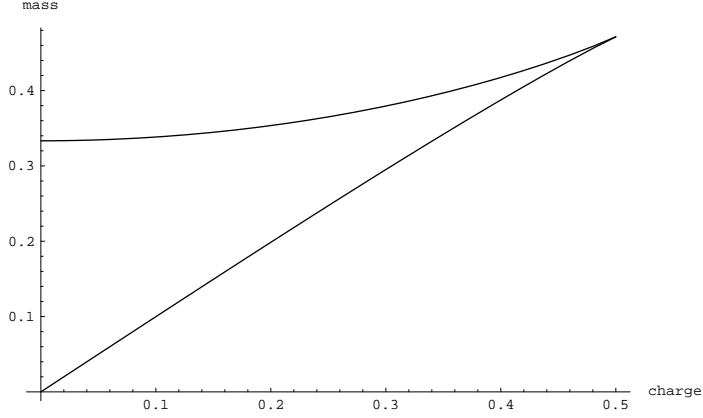,height=3in}}
\vspace*{-4ex}
\caption{Allowed values of mass $\mu$ and charge $g$.}
\label{fig:mvsg}
\end{figure}
The curve $\mu_{min}(g)$ corresponds to $r_{-} = r_{+}$,
i.e., coincident inner and outer horizons, while the curve
$\mu_{max}(g)$ corresponds to $r_{+}=r_c$. These two conditions can be
combined into a simple form $g^2=\rho^2 (1-\rho^2)$. 

The de Sitter black hole action is thus a
function of four variables, $\mu$, $g$, $b \equiv \beta \sqrt{\Lambda}$,
and the box size $R \equiv r_B \sqrt{\Lambda}$. To locate a saddle point,
we must find the roots of $V(r)=0$, choose the middle positive
root, substitute it into the action, and minimize it, subject to the
condition that $\mu$ and $g$ lie in the interior of the  
region depicted in Fig.\ \ref{fig:mvsg}. 

To analyze the stability of the black hole, it is convenient to revert to the
variables $r_+$ and $e$ rather than their ``normalized'' versions $\mu$
and $g$. We will also stay away the extremal black holes
corresponding to the curves $\mu_{min}(g)$ and $\mu_{max}(g)$, by
requiring that $g^2<\rho_+^2(1-\rho_+^2)$ and $g^2 <
\rho_c^2(1-\rho_c^2)$.
Requiring that the $r$-$\tau$ plane look like ${\mathbb R}^2$ near
$r=r_+$ fixes $C$ to be $r_+ + e^2/r_+ - \Lambda r_+^3/3$, which puts
$V(r)$ in the form
\begin{equation} 
V(r)=\left(1-\frac{r_+}{r}\right)\left(1- \frac{e^2/r_+}{r} -
\frac{\Lambda}{3}(r^2 + r r_+ + r_+^2) \right) .
\end{equation}
Thermodynamic quantities are specified at a boundary $r=r_B$, 
and we require that $r_B < r_c$. For any given value of
$r_+$ and $e$, the maximum allowed value of $r_B$ is given by the
larger root of the equation $g^2=\rho^2 (1-\rho^2)$.  
To summarize, for any given value of $x=r_+/r_B$, the allowed values
of charge $q=e/r_B$ are given by $q^2<x^2(1-R^2 x^2)$ and $q^2<1-R^2$.

Subject to the above conditions, the reduced action can be shown to be
\begin{eqnarray} 
I_C&=&\beta r_B [1-\sqrt{V(r_B)}]-\pi r_{+}^2 \\ 
&=& \beta r_B \left[1-\sqrt{\left(1-\frac{r_+}{r_B}\right)
\left(1-\frac{e^2}{r_B r_+} - \frac{\Lambda}{3} (r_B^2 + r_B r_+ +
r_+^2)\right)}\right] -\pi r_+^2 .
\label{dSaction}
\end{eqnarray} 
Extremizing that action with respect to $x$ and solving for $b$, we obtain
\begin{equation} 
\beta = \frac{4 \pi r_+ \sqrt{(1 - \frac{r_+}{r_B})
  \left(1-\frac{e^2}{r_+ r_B} -\frac{\Lambda}{3} (r_B^2 + r_+ r_B +
  r_+^2)\right)}}{1-\frac{e^2}{r_+ r_B} - \frac{\Lambda}{3}(r_B^2 +
  r_+ r_B + r_+^2) + \left(1-\frac{r_+}{r_B}\right) \left(
  \frac{\Lambda}{3} (r_B^2 + 2r_+ r_B) - \frac{e^2}{r_+^2}\right)} .
\label{dstemp}
\end{equation} 
This is a ninth order algebraic equation for $r_+$, which may be 
written as
\begin{equation} 
x^5 (1-x) \left(x - q^2 - \frac{R^2 x (1+x+x^2)}{3}\right) - b^2 (q^2
- x^2 + R^2 x^4)^2 = 0.
\label{ninth}
\end{equation} 

\begin{figure}
\begin{tabular}{ll} 
\epsfig{figure=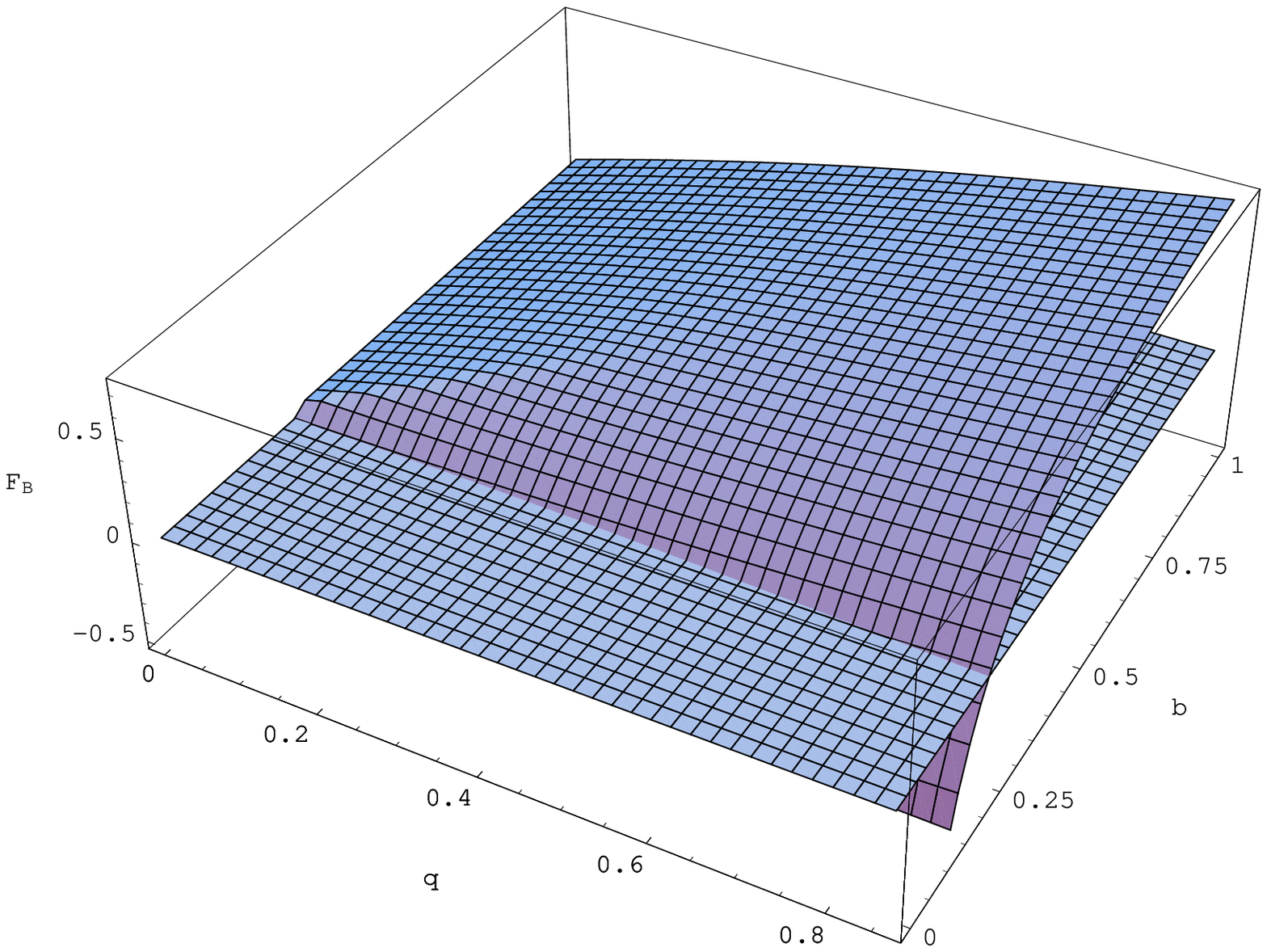,width=.45\textwidth} &
\epsfig{figure=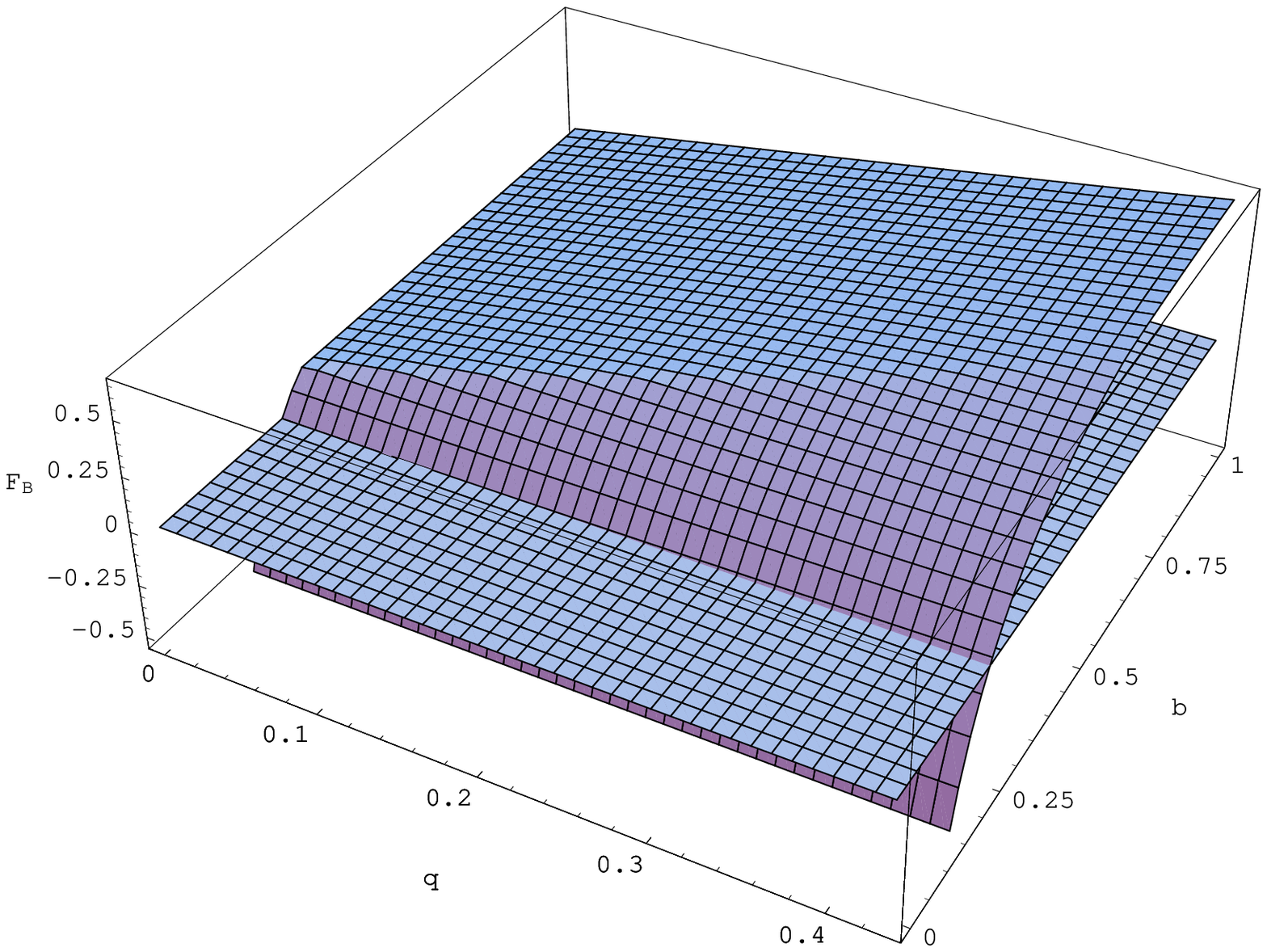,width=.45\textwidth} \\[1ex]
\small\hspace*{6em}(a) $R=0.5$ & \small\hspace*{6em} (b) $R=0.9$ \\  
\end{tabular}
\caption{ Helmholtz free energy of the black hole for two values of $R$.}
\label{ds}
\end{figure}

Given values for $b,q$ and $R$, (\ref{ninth}) can be solved
numerically for $x$.  In Figure \ref{ds}, we plot the free energy as a
function of charge $q$ and inverse temperature $b$ for two values
of the cavity radius $R$.  As before, we also show the $F=0$ plane,  
which should give an approximate locus for the
Hawking-Page phase transition.  

\begin{figure}
\begin{tabular}{ll} 
\epsfig{figure=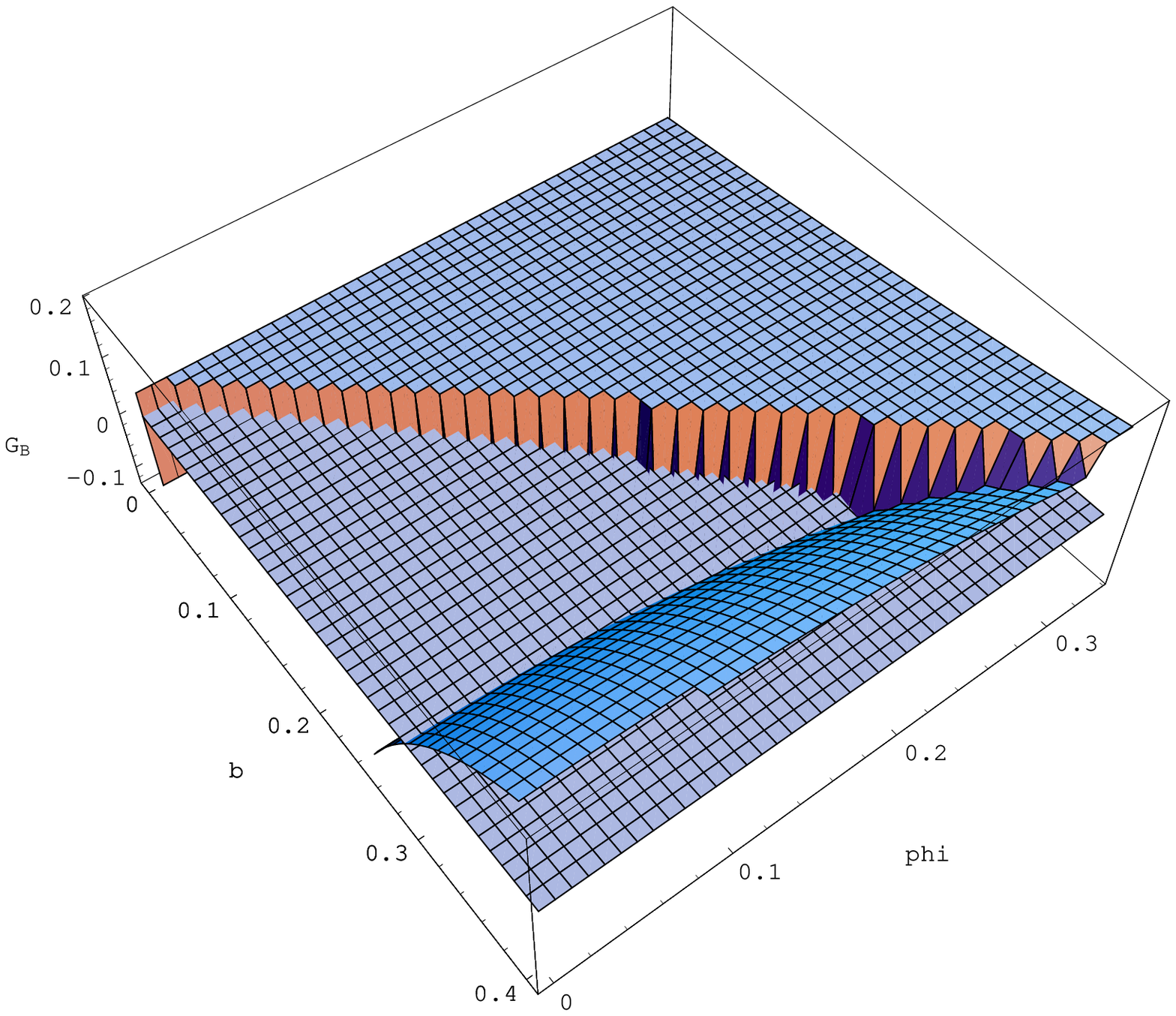,width=.45\textwidth} &
\epsfig{figure=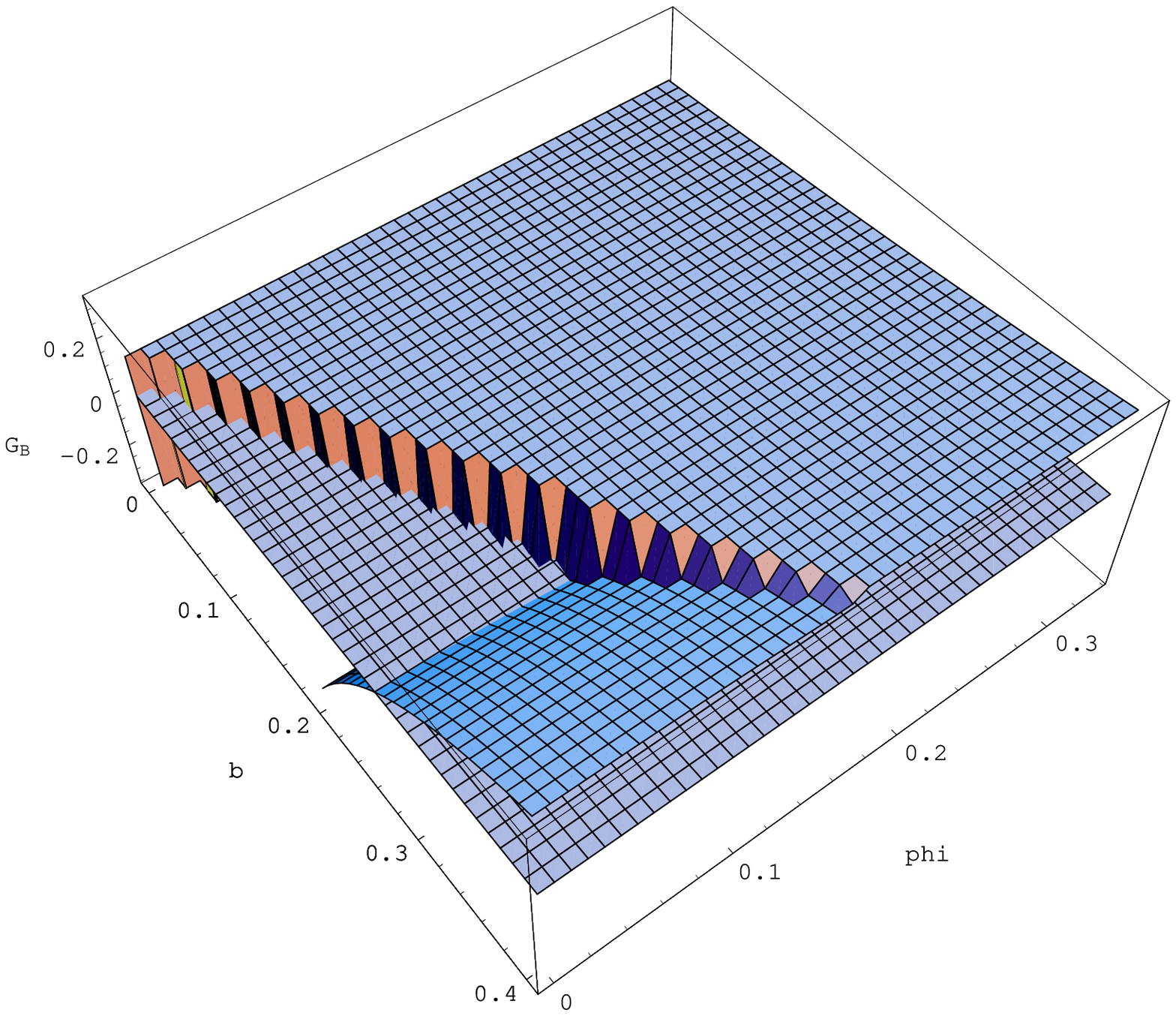,width=.45\textwidth} \\[3ex]
\small\hspace*{6em}(a) $R=0.3$ & \small\hspace*{6em} (b) $R=0.5$ \\ 
\end{tabular}
\caption{Gibbs free energy of the black hole 
for two values of $R$.}
\label{dsg}
\end{figure}

As in the asymptotically flat
case, we have checked this behavior in the grand canonical ensemble.
Figure \ref{dsg}
shows the Gibbs free energy and the $G=0$ plane, and confirms the
existence of a phase transition.  Note that a Hawking-Page phase 
transition occurs for all values of $\Lambda$.  While the presence 
of a cosmological constant affects the transition temperature, it does 
not change the qualitative behavior.

For small values of $R$, it is clear that the charged black hole in de Sitter 
space behaves very much like the corresponding black hole in flat space.   
There is again a line of first order phase transitions in the supercooled 
region that ends at a critical point.  For large values of $R$, the critical 
point disappears: the place at which it would occur is essentially pushed 
outside the cosmological horizon.

In general, we have been unable to find the nature of the phase
transition at the critical point.  We can, however, do so for for two important 
special cases: the uncharged black hole $(e=0)$ and the case of small
cosmological constant.  For $e=0$, the function $b$ takes the form
\begin{equation} 
b(x,R)=\frac{x \sqrt{(1-x)(1-R^2(1+x+x^2)/3)}}{1-R^2 x^2} .
\end{equation}  
The positive root of the equation $1-\Lambda(r^2 + r_+ r+ r_+^2)/3=0$
gives us the location of the cosmological horizon $r_c$. Requiring
that $r_+ < r_c$ tells us that $r_+^2 < 1/\Lambda$. Similarly, $r_+
<r_B < r_c$ implies that $R\equiv r_B \sqrt{\Lambda} <1$.

\begin{figure} 
\centerline{\epsfig{figure=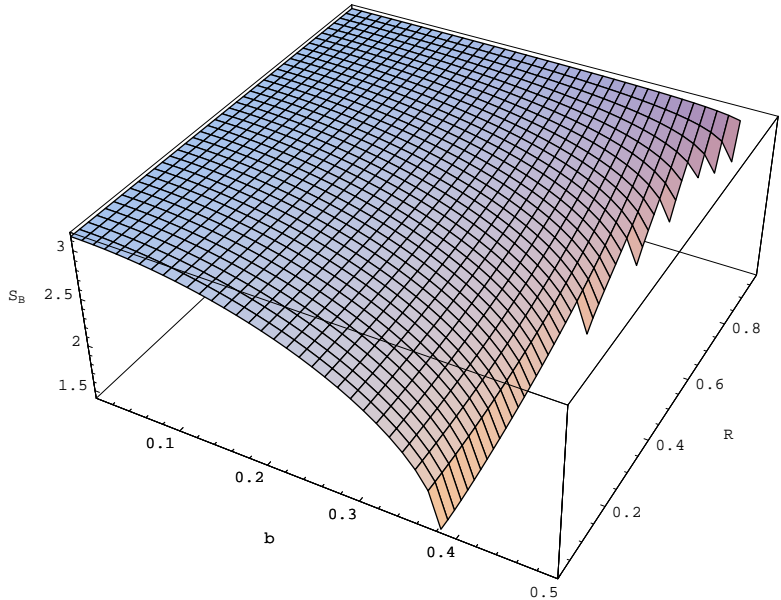,height=3.2in}}
\caption{Free energy for the uncharged de Sitter black hole as a function 
  of inverse temperature $b$ and cosmological constant $R=\sqrt{\Lambda}r_B$.}
\label{fig:svsR0}
\end{figure}

The free energy  can be plotted as a function of $b$ and $R$
(Figure \ref{fig:svsR0}), and clearly shows that black holes do not
exist above a critical value $b_c$ of $b$, for any fixed value of $R$. 
This critical value is analogous to the value discussed at the end of
section \ref{HP1}: it is a limiting temperature beyond which not even
a supercooled black hole can exist. 
For any given value of $R$, the entropy of the system behaves as
$S-S_c(R) \sim (b-b_c(R))^{\alpha}$. By choosing various values of
$R$, one can check that $\alpha=1/2$ and that $C_v \sim
(\tau-\tau_c)^{-1/2}$. The critical behavior thus is very much like
that of the uncharged black hole in flat space discussed in section \ref{Unch}.

For the case of small $\Lambda$, we can expand $b$ in powers of $R^2$
and look for critical points in the supercooled regime. Both $b_c$ and $q_c$ 
receive correction to order $R^2$, while $x_c$ remains the same to this order. 
The critical exponents also remain the same.

\section{Discussion}

The thermodynamics of black holes is interesting in its own right, and also
because of its connections to holography and the AdS/CFT correspondence. 
Largely inspired by the latter connection, past investigations have found a
rich phase structure for asymptotically anti-de Sitter black holes.  

Our main result is that this phase structure does not, in fact, require 
asymptotically anti-de Sitter boundary conditions, but appears as well 
for black holes in cavities in asymptotically flat and asymptotically de Sitter 
spaces.  In particular, both Hawking-Page-like phase transitions and 
transitions between small and large black holes occur as we  vary the 
charge and temperature.  For ``supercooled'' black holes in flat space, 
and black holes in de Sitter space with small $\Lambda$, we found a line 
of first order phase transitions ending at a second order transition, and
we were able to identify the critical exponents at the second order point.
Remarkably, these were the same as those for asymptotically 
anti-de Sitter black holes, despite a very different algebraic structure 
of the partition function.

For anti-de Sitter space, the phase structure of charged black holes has
served as a valuable test of the AdS/CFT correspondence.  We leave open
the very interesting question of whether the corresponding results for
asymptotically flat and asymptotically de Sitter black holes might cast
light on proposals for finite-volume holography in this more general
setting.

{\bf Acknowledgments} This work is supported in part by DOE grant 
DE-FG03-91ER40674.

\end{document}